\documentclass[fleqn,usenatbib]{mnras}

\usepackage{newtxtext,newtxmath}

\usepackage[T1]{fontenc}
\usepackage{ae,aecompl}

\usepackage[nointegrals]{wasysym}
\usepackage[dvipsnames]{xcolor}
\usepackage{hyperref}
\hypersetup{
  colorlinks,
  citecolor=Aquamarine,
  linkcolor=red
}

\def\aap{\ {A\&A}\ }

\def\aj{\ {AJ}\ }

\def\apj{\ {ApJ}\ }
\def\apjl{\ {ApJL}\ }

\def\apss{\ {Ap\&SS}\ }
\def\araa{\ {ARA\&A}\ }

\def\mnras{\ {MNRAS}\ }

\def\prd{\ {Phys. Rev. D.}\ }

\def\ssr{\ {Space Sci. Rev.}\ }

\input{changes.tex}

\newcommand{\icnu}{IceCube-190331A}
\newcommand{\wisea}{WISEA\,J222925.59$-$201846.0}
\newcommand{\wga}{1WGA\,J2229.4$-$2018}
\newcommand{\mass}{2MASS\,J22292559$-$20184}

\newcommand{\swift}{\textsl{Swift}}
\newcommand{\nustar}{\textsl{NuSTAR}}
\newcommand{\xshooter}{\mbox{X-shooter}}
\newcommand{\fermi}{\textsl{Fermi}}

\newcommand{\xray}{\mbox{X-ray}}

\newcommand{\gray}{\mbox{$\gamma$-ray}}

\def\arcmin{\hbox{$^\prime$}}
\def\arcsec{\hbox{$^{\prime\prime}$}}
\def\farcs{\hbox{$.\!\!^{\prime\prime}$}}

\newcommand{\eqb}{\begin{equation}}
\newcommand{\eqe}{\end{equation}}

\def\gtrsim{\ {\raise-.5ex\hbox{$\buildrel>\over\sim$}}\ }
\def\lesssim{\ {\raise-.5ex\hbox{$\buildrel<\over\sim$}}\ }
\def\simlt{\mathrel{\hbox{\rlap{\hbox{\lower4pt\hbox{$\sim$}}}\hbox{$< $}}}}
\def\simgt{\mathrel{\hbox{\rlap{\hbox{\lower4pt\hbox{$\sim$}}}\hbox{$> $}}}}

\newcommand{\unsim}{\mathord{\sim}}

\usepackage{graphicx}	
\usepackage{amsmath}	

\title[Multiwavelength counterparts to IceCube-190331A]{Multimessenger observations of counterparts to IceCube-190331A}

\author[F. Krau\ss{} et al.] {%
Felicia Krau\ss{},$^{1}$\thanks{E-mail: Felicia.Krauss@psu.edu}
Emily Calamari,$^{2}$\thanks{ecalamari@gradcenter.cuny.edu}
Azadeh Keivani,$^{3,4}$\thanks{azadeh.keivani@columbia.edu}
Alexis Coleiro,$^{5}$
\newauthor
Phil A. Evans,$^{6}$
Derek B. Fox,$^{1,7,8}$
Jamie A. Kennea,$^{1}$
Peter M\'esz\'aros,$^{1,7,8}$
\newauthor
Kohta Murase,$^{7}$
Thomas D. Russell,$^{9}$
Marcos Santander,$^{10}$
and Aaron Tohuvavohu$^{1,11}$
\vspace*{0.25cm}\\
$^{1}$Department of Astronomy \& Astrophysics, Pennsylvania State University, University Park, PA 16802, USA\\
$^{2}$Department of Physics, Barnard College, Columbia University, New York, NY 10027, USA\\
$^{3}$Department of Physics, Columbia University, New York, NY 10027,
USA\\
$^{4}$Columbia Astrophysics Laboratory, Columbia University, New York, NY 10027, USA\\
$^{5}$APC, Univ Paris Diderot, CNRS/IN2P3, CEA/Irfu, Obs de Paris, Sorbonne Paris Cit\'e, France\\
$^{6}$Department of Physics \& Astronomy, University of Leicester,
Leicester, LEI 7RH, UK\\
$^{7}$Center for Particle \& Gravitational Astrophysics, Institute for
Gravitation and the Cosmos, Pennsylvania State University, University~Park, PA 16802, USA\\
$^{8}$Center for Theoretical \& Observational Cosmology, Institute for
Gravitation and the Cosmos, Pennsylvania State University, University
Park, PA 16802, USA\\
$^{9}$Anton Pannekoek Institute for Astronomy, University of
Amsterdam, Science Park 904, 1098XH Amsterdam, the Netherlands\\
$^{10}$Department of Physics and Astronomy, University of Alabama,
Tuscaloosa, AL 35487, USA\\
$^{11}$University of Toronto, Toronto, Canada\\
}

\date{Accepted 2020 July 17. Received 2020 July 3; in original form 2020 May 22}

\pubyear{2020}

\begin{document}
\label{firstpage}
\pagerange{\pageref{firstpage}--\pageref{lastpage}}
\maketitle

\begin{abstract}
 High-energy neutrinos are a promising tool for identifying
 astrophysical sources of high and ultra-high energy cosmic rays
 (UHECR). Prospects of detecting neutrinos at high energies
 ($\gtrsim$TeV)  from blazars have been boosted after the recent
 association of IceCube-170922A and TXS\,0506+056. We investigate the
 high-energy neutrino, IceCube-190331A, a high-energy starting event
 (HESE) with a high likelihood of being astrophysical in origin. We
 initiated a \swift/XRT and UVOT tiling  mosaic of the neutrino
 localisation, and followed up with ATCA radio observations, compiling
 a  multiwavelength SED for the most likely source of origin.
 \nustar\ observations of the neutrino location and a nearby X-ray
 source were also performed. We find two promising counterpart in the
 90\% confidence localisation region and identify the brightest as the
 most likely counterpart. However, no \fermi/LAT $\gamma$-ray source
 and no prompt \swift/BAT source is consistent with the neutrino
 event. At this point it is unclear whether any of the counterparts
 produced IceCube-190331A. We note  that the Helix Nebula is also
 consistent with the position of the neutrino event, and we calculate
 that associated particle acceleration processes cannot produce the
 required energies to generate a high-energy HESE neutrino. 
\end{abstract}

\begin{keywords}
neutrinos --- galaxies: active --- BL Lacertae objects:
  general --- quasars: general --- galaxies: jets
\end{keywords}



\section{Introduction}
\label{sec:intro}
Cosmic rays arriving at Earth have been detected up to the extreme
energies of $10^{21}$\,eV since more than a century ago -- yet their
origin remains elusive \citep[e.g.,][]{Norman1995}.  
A promising tool for identifying the astrophysical sources of high
and ultra-high energy cosmic rays are high-energy neutrinos as they
are not deflected in interstellar and intergalactic magnetic fields.  
Consistent with these expectations, a diffuse extraterrestrial flux of
high-energy neutrinos has been observed by the IceCube neutrino
observatory over more than a decade of
observation~\citep{Aartsen:2013jdh,PRL2014,chaack2017}. Localisation
uncertainties mean that the nature of these neutrinos is still
unknown.

Blazars and other types of AGN have been predicted to produce neutrinos in jets
\citep{Biermann1987,Mannheim1991,mannheim_1993,mannheim_1995} and/or
in their cores
\citep{Eichler1979,Berezinskii1981,Begelman1990,Stecker1991,Stecker2013}.
TeV and PeV neutrinos are expected from flat-spectrum radio quasars,
while BL Lacs are expected to produce neutrinos at EeV energies.
Recently, progress has also been made in explaining neutrinos from BL
Lac objects \citep{Murase:2014foa,Dermer:2014vaa,Tavecchio2015}.  We
have shown that blazars can calorimetrically explain IceCube neutrinos
\citep{Krauss2014}, low-significance
coincidence between a blazar outburst and an astrophysical neutrino,
IC\,35 \citep[and PKS\,1424$-$418;][]{bb}. To date, the neutrino
candidate, IceCube-170922A has only been the second $\gtrsim$3$\sigma$
association of neutrino emission to an astronomical source
\citep[SN\,1987A; blazar TXS~0506+056;][]{ic170922a, keivani2018}.
However, it has also been shown that for realistic neutrino spectra,
blazars can account for all IceCube high-energy neutrinos
\citep{Krauss2018}; this is in disagreement with the 30\% limit on the
contribution of blazars found by \citet{IceCube2017} for all IceCube
neutrinos. More stringent limits of 5--15\% have been placed by
\citet{Hooper2019,Yuan2019}, which would still be consistent with a
significant contribution of AGN, including blazars, to the PeV neutrinos
\citep{Murase2016}. It is possible that non-blazar AGN produce the
entire or a large fraction of the astrophysical neutrino flux seen by
IceCube \citep{Hooper2019}. 
Some authors have argued for a combination of BL Lac and pulsar wind
nebula as the origin of the IceCube neutrinos \citep{Padovani2014}.

While blazars and other active galactic nuclei (AGN) are excellent
candidates for accelerating cosmic rays to ultra-high energies,
significant contributions from other types of sources are not yet
ruled out. Suggested populations include starburst galaxies
\citep{Murase2013,smm+15,Bechtol2017}, and (choked) GRBs
\citep{Murase2013b,Senno2016,Tamborra2016,Aartsen2017}.

IceCube and the Astrophysical Multimessenger Observatory
Network (AMON)\footnote{see \url{https://www.amon.psu.edu/} for
  details.} started a real-time program in 2016~\citep{IC_realtime} to
identify and localise high-energy neutrinos in order to distribute
them to follow-up observatories. Since 2019 IceCube provides these
alerts at ``bronze'', ``silver'' and ``gold'' levels. There have been
17 such alerts as of February 2020 (10 ``bronze'' alerts and 7
``gold'' alerts), several of which resulted in extensive
multimessenger campaigns to observe the location of the neutrino
candidate in different wavelengths/messengers. IceCube-170922A has so
far been the only event with a $\gtrsim 3\sigma$ source identification
\citep{ic170922a}.

On March 31, 2019, the IceCube Neutrino Observatory identified a high
energy neutrino candidate (labelled IceCube-190331A), likely produced
by a muon neutrino. This event was publicly distributed 
through the gamma-ray coordinates network (GCN;~\citealt{gcn95})
within 34 seconds~\citep{GCN_notice_icnu}. 
A subsequent search by \fermi/LAT determined there were no known
\gray\ sources within the 90\% \icnu\ localisation
error~\citep{gcn_fermi}.
Given the event direction, this paper seeks to investigate a possible
origin for this high-energy neutrino candidate by conducting follow-up
observations of known sources within the uncertainty region. 

We report the multiwavelength observations (radio, optical, UV, X-ray)
of possible counterparts detected by \swift/XRT during follow-up observations
of the \icnu\ high-energy neutrino candidate (Sec.~\ref{sec:obs}) and
discuss possible associations with the IceCube event (Sec.~\ref{sec:disc}).

Throughout  the  paper  we  use  the  standard  cosmological model
with $\Omega_\mathrm{m}=0.3$, $\Lambda=0.7$,
$H_0=70$\,km\,s$^{-1}$\,Mpc$^{-1}$ \citep{Beringer2012}.

\section{Observations}
\label{sec:obs}

In this section, we review the detection of the \icnu\ neutrino event,
and present observations and data analysis of electromagnetic
follow-up observations from  \swift\ (\xray, ultraviolet/optical,
\gray), ATCA (radio), \xshooter\ (ultraviolet, optical,
near-infrared), and \nustar\ (\xray).\\

\subsection{IceCube Detection}
\label{sub:obs:icecube}
On March 31, 2019, the IceCube Neutrino Observatory identified a
high-energy neutrino candidate through its High Energy Starting Event
(HESE) stream. This event had a high probability of being produced by
a muon neutrino of astrophysical origin with a deposited charge of
about 198736.44 photoelectrons in the detector
~\citep{GCN_notice_icnu}.  This event was recorded as having one of
the highest deposited energies ever seen, making it a promising
astrophysical neutrino candidate~\citep{GCN_icnu}. After conducting
a ground-based analysis using offline reconstruction algorithms, IceCube
was able to report an event direction at
RA=$337\fdg68^{+0.23}_{-0.34}$, Dec=$-20\fdg70^{+0.30}_{-0.48}$
~\citep[J2000; 90\% containment ellipse;][]{GCN_icnu,atel1903}.
Subsequently, an additional search for track-like muon neutrino events
arriving from the direction of \icnu\ was performed by IceCube for two
days after the initial event time, as well as a search to include the
previous month of data. No additional track-like events were found
within the 90\% spatial containment region in either search
\citep{GCN_icnu_follow}. 

Several multiwavelength follow-up observations were conducted in order
to find potential EM counterparts to the very high energy neutrino
candidate. Although these searches did not find any high-confidence EM
counterpart \citep{gcn_fermi}, we discuss two possible
counterparts below (see Sect.~\ref{sec:disc}). 

\subsection{\swift/BAT prompt observations}
At the time of arrival of \icnu\ (T0), the neutrino localisation
region was serendiptidously located near the highest sensitivity
location of the coded field-of-view (89\% partial coding fraction) of
the \swift\ Burst Alert Telescope \citep[BAT;][]{swift:bat}. This
allows us to set sensitive upper limits on the existence of a prompt
gamma-ray transient coincident with (or directly preceding or
succeeding) the high-energy neutrino emission. We perform a blind
search on the BAT raw light curves with time-bins of 64 ms, 1 s, and
1.6 s. We find no evidence for any short or long GRB-like emission
within T0 $\pm$ 500 s of the neutrino arrival time, and set a
conservative 5$\sigma$ upper limit for any short GRB of
$\unsim 1.5\times10^{-7}$\,erg\,$\mathrm{s}^{-1}\mathrm{cm}^{-2}.$ 

We also performed a search for longer time-scale emission, on a survey
image produced by the BAT from T0-580s to T0+ 660s. We find no new or
uncatalogued hard X-ray sources within the neutrino localisation
region, and set a 3$\sigma$ flux upper limit of $\sim$15 mCrab assuming a
powerlaw spectral index of $\Gamma = 2.15$.

\subsection{\swift/XRT observations}
\label{sub:obs:swift-xrt}

\begin{figure}
  \centering
  \includegraphics[width=0.95\columnwidth]{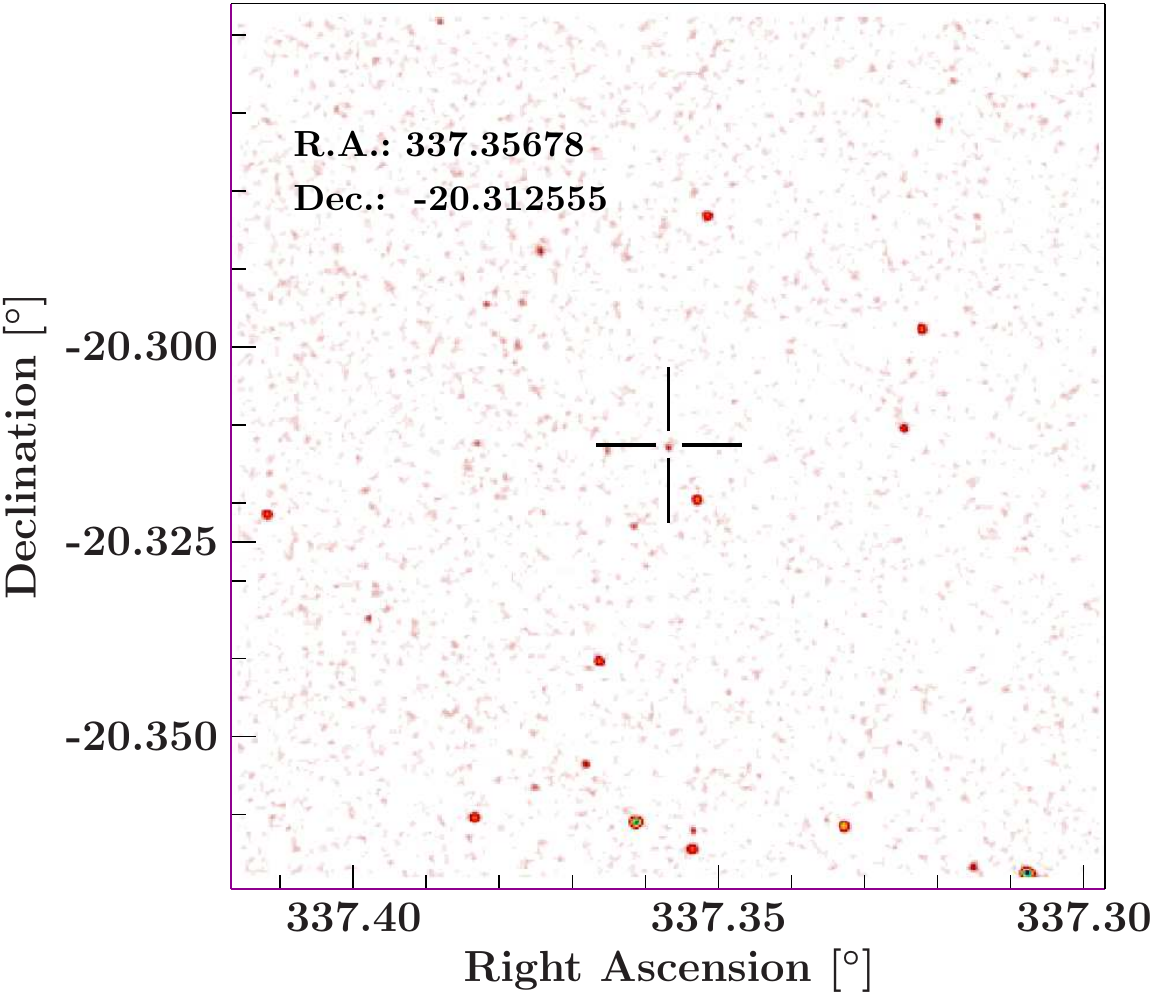}
  \caption{K-Band 2MASS image of the region surrounding
    2MASS\,J22292559$-$2018462.}
  \label{fig-2mass}
\end{figure}

\icnu\ triggered the Neil Gehrels \swift\ Observatory in automated
fashion via the AMON cyberinfrastructure~\citep{amon19}, however
prompt observation of the neutrino localisation was not possible with
\swift\ as it was initially within the satellite's Sun avoidance region.
\begin{table}
  \caption{Source found by \swift/XRT with their best fit centroid
    position, as well as source significance. Source \#4 is at the
    lowest significance.}
  \label{tab-srcs}
  \begin{tabular}{lccc}
    Sources & R.A. & Dec. & Significance\\
       ID  & [J2000] & [J2000] & $\sigma$ \\
    \hline
    1 & 337.3551 & $-$20.31325 & 5.56 \\
    2 & 337.5285 & $-$21.0994 & 5.06\\
    3 & 338.0251 & $-$21.0493 & 4.52\\
    4 & 338.0184 & $-$21.1199 & 4.16 \\
    \end{tabular}
\end{table}
\swift\ observations of the \icnu\ field began on April 9, 2019, nine days
after the event. \swift\ was able to observe a region of
approximately 33\arcmin\ radius centred on the event direction of
RA=$337\fdg68$, Dec=$-20\fdg70$ (J2000), using an on-board 7-point
tiling pattern \citep{gcn_swift}. During this initial observation,
\swift/XRT collected approximately $800$ seconds of data per tile for
a total of $\unsim 5540$ seconds the morning of April 9 and was able to
detect four \xray\ sources (see Table~\ref{tab-srcs}) with the new XRT
detection system \citep{Evans2019}. The highest significance (source
\#1) found during these observations is catalogued at
RA=$337\fdg35513$, Dec=$-20\fdg31324$ (J2000; 90 \% containment 
region), matching the known \xray\ source \wga\ from ROSAT/WGACAT catalogue
\citep[15.37\arcsec\ distance; ][]{wga}. Another source,
5.3\arcsec\ away from \xray\ source \# 1 is listed in the Milliquas catalogue
\citep{flesch2017} as the likely AGN  \wisea\ and is consistent with
the J=16.74\,mag 2MASS source \citep[2MASS\,J22292559$-$2018462, see
  Fig.~\ref{fig-2mass};][]{mass2,gcn_swift}. \wisea\ is not listed in
the most recent ALLWISE catalogue, so the WISEA detection might not be
real \citep{wise}.

Although it seems that the sources (\wisea\ /\mass\ and \wga ) are the
same, it is possible that neither \wga\ nor \wisea\ are real source
detections, but rather background fluctuations, given their low
detection significance and distance from the \mass\ coordinates.

\swift/XRT performed a further observation centred on the location of
source \#1.
The \xray\ analysis following the observation period $58582.323$ -
$58582.471$ MJD  focused on data from the most
significant sources and did not consider the lower-significance
\xray\  sources. The data used in analysis were from \swift/XRT
observations of the position 
of \wga\ (source \#1) on April 9, 2019 and April 16, 2019. The newest calibration
was applied using the xrtpipeline to the raw data using HEASoft (V. 6.26).
Spectra were extracted from the reprocessed image using a source region
with 54.218\arcsec radius and an annulus for the background
region of 82.506 and 235.721\arcsec using XSELECT (V. 2.4). Due to the
low count rate, data were binned to a signal-to-noise-ratio (SNR) of 1
in the \textsl{Interactive Spectral Interpretation System}
\citep[\textsl{ISIS}; V. 1.6.2-44;][]{isis}. Due to the low SNR we use
Cash statistics to find a best fit \citep{cash}. The data were fit
with an absorbed power law with a convolution model to calculate the 
flux. For the absorption model we use tbnew\footnote{online at:
  \url{http://pulsar.sternwarte.uni-erlangen.de/wilms/research/tbabs/}}
with the \texttt{vern} cross-sections \citep{Verner1996} and the
\texttt{wilm} abundances \citep{Wilms2000}.  The hydrogen equivalent
absorption column density was frozen to a value of
$3.55\times10^{20}$\,$\text{cm}^{-2}$ \citep{h14pi}. 
\begin{figure*}\centering
  \includegraphics[width=0.95\textwidth]{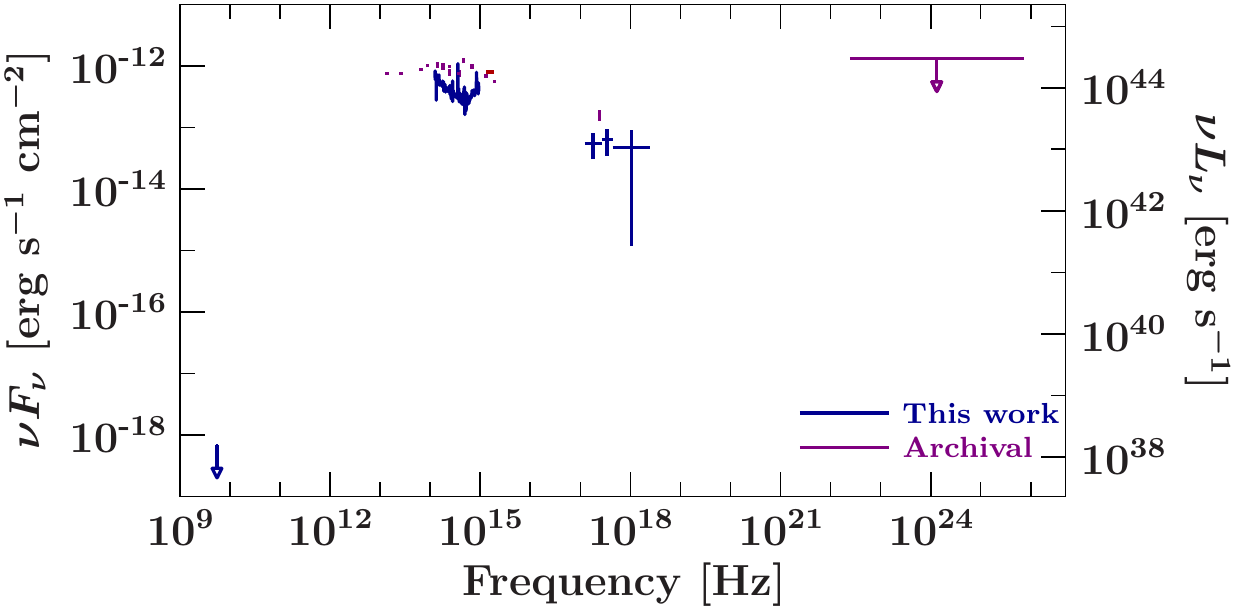}
  \caption{Multiwavelength SED of source \#1, including our ATCA,
    X-Shooter, and \swift /XRT data. Archival data is shown in purple.}
  \label{fig-sed}
\end{figure*}

The photon indices for the observations on April 9 and April 16 were
found to be  $\Gamma_1 = 2.4^{+4.3}_{-1.4}$ and $\Gamma_2=2.4^{+1.6}_{-0.8}$,
respectively, with corresponding flux values of
$(1.1^{+0.7}_{-0.9})\times10^{-13}$ and
$(1.6^{+1.0}_{-0.9})\times10^{-14}$\,erg\,s$^{-1}$\,cm$^{-2}$,
indicating a possible change in flux.  
Uncertainties for both the photon index and flux were calculated at the 90\%
confidence level.
Both observations are combined in the SED and shown with a
signal-to-noise ratio binning of 2 (see Fig.~\ref{fig-sed}).

\subsection{\swift/UVOT follow-up observations}
\begin{figure}
    \centering
    \includegraphics[width=\columnwidth]{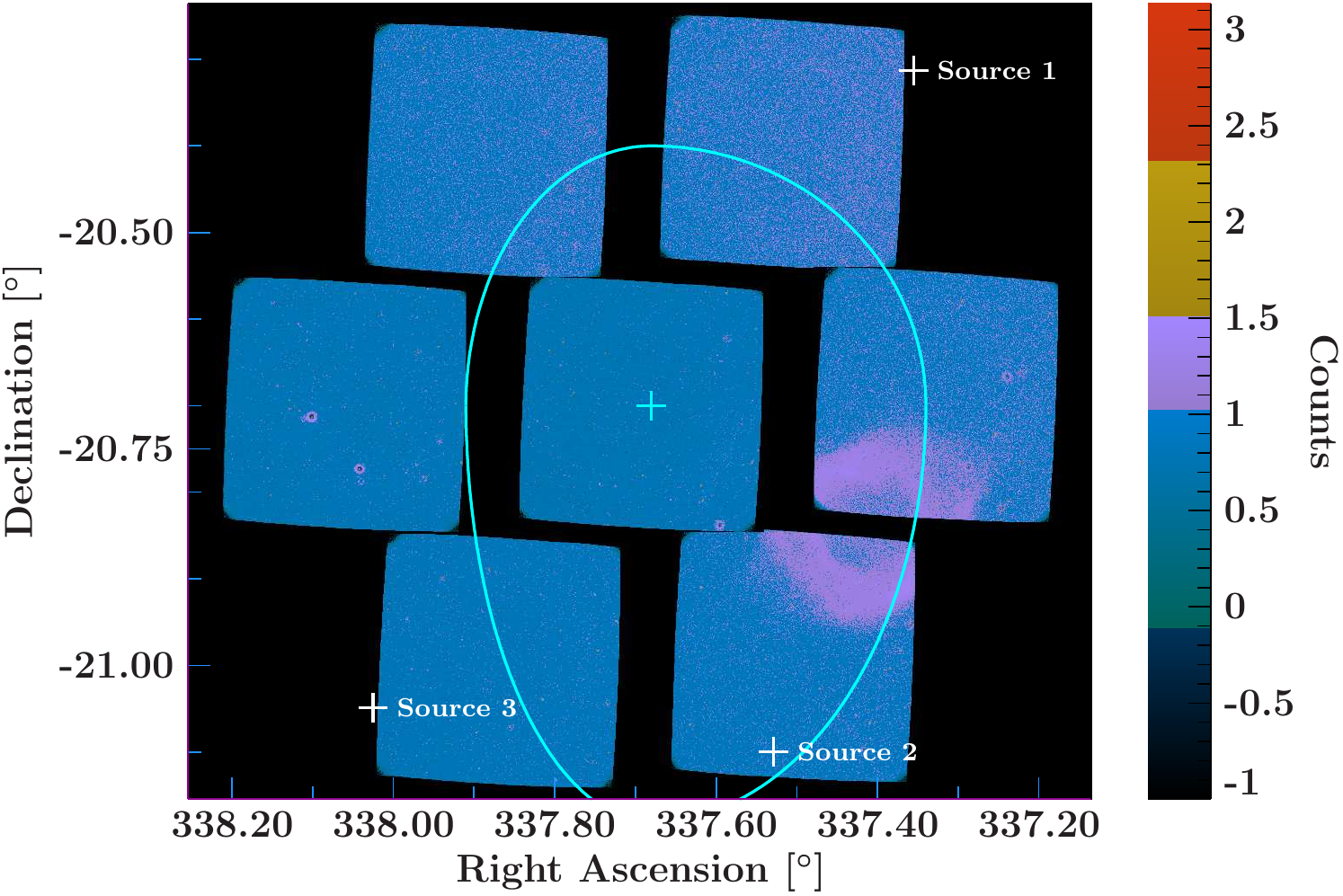}
    \caption{\swift/UVOT mosaic follow-up of the localisation region
      of \icnu. The Helix nebula, NGC 7293,  is clearly visible in the
      west. It is not detected at X-ray energies. The cyan cross shows
    the best fit position of the neutrino event. The 90\% containment
    region is given in cyan, following the values from  \citet{atel1903}.}
    \label{fig:uvot}
\end{figure}

The \swift/UltraViolet-Optical Telescope (UVOT) \citep{uvot2005} also
participated in both the tiling and targeted follow-up in response to
\icnu. Only source \#2 was within the field of view of the initial
tiled UVOT observations (see Fig.~\ref{fig:uvot}). An additional
follow-up observation on April 16 provided a UVOT image for source
\#1 for a cleaned exposure time of 2829\,s in the \textit{uvw2} filter. UVOT
images in one observations were summed using uvotimsum. Source counts
were extracted using uvot2pha with a 5\arcsec region at an updated
source position of RA=337$\fdg$5298451, Dec=$-21\fdg09967127$ and
annulus for the background region of 13 and 26\arcsec, centred on
the source position while ensuring no contamination from background
sources.
Source \#1 was extracted with regions of the same size, centred on
RA=337$\fdg$3566258, Dec=$-20\fdg3127867$. 
Additionally, \textit{uvw2} images detected WISEA J222925.59-2201846.0
at an AB magnitude of $19.64^{\pm 0.08 \mathrm{(stat)}}_{\pm 0.03
  \mathrm{(sys)}}$. No flux variability or changes were detected in
the four observations.

We performed a search for uncatalogued sources in the UVOT
\textit{u}-band with observations taken during the 7-point tiling
follow-up. No new or uncatalogued sources were found down to an
average 5$\sigma$ upper limit of \textit{u}=20.3 mag AB.

\subsection{X-Shooter Observation}
\label{sub:obs:xshooter}
Medium-resolution spectroscopy of 1WGA J2229.4-2018 was obtained with
the X-shooter spectrograph~\citep{Vernet2011} of the Very Large
Telescope (VLT) UT2 at the ESO Paranal Observatory on 2019 April 26.
The three arms of X-shooter, (UV: UVB, optical: VIS and 
near-infrared: NIR) were used with slit widths of 1\farcs0,
0\farcs9, and 0\farcs9, respectively. These data provide
quasi-simultaneous 300--2400\,nm spectral coverage with average
spectral resolutions $\lambda/\Delta\lambda$ of 5400, 8900, and 5600,
respectively, in each arm. Observing conditions were intermediate,
with a clear sky, a seeing of $\sim$1\farcs5, and an airmass of 1.7.
Individual exposure times are 445 s, 352 s, and 200 s for the UBV,
VIS, and NIR arms, respectively, which lead to a total integration
times of 3560 s, 2816 s, and 3200 s. Standard ABBA nodding observing
mode was used to allow for an effective background subtraction.
\begin{figure*}
    \centering
    \includegraphics[width=0.8\textwidth]{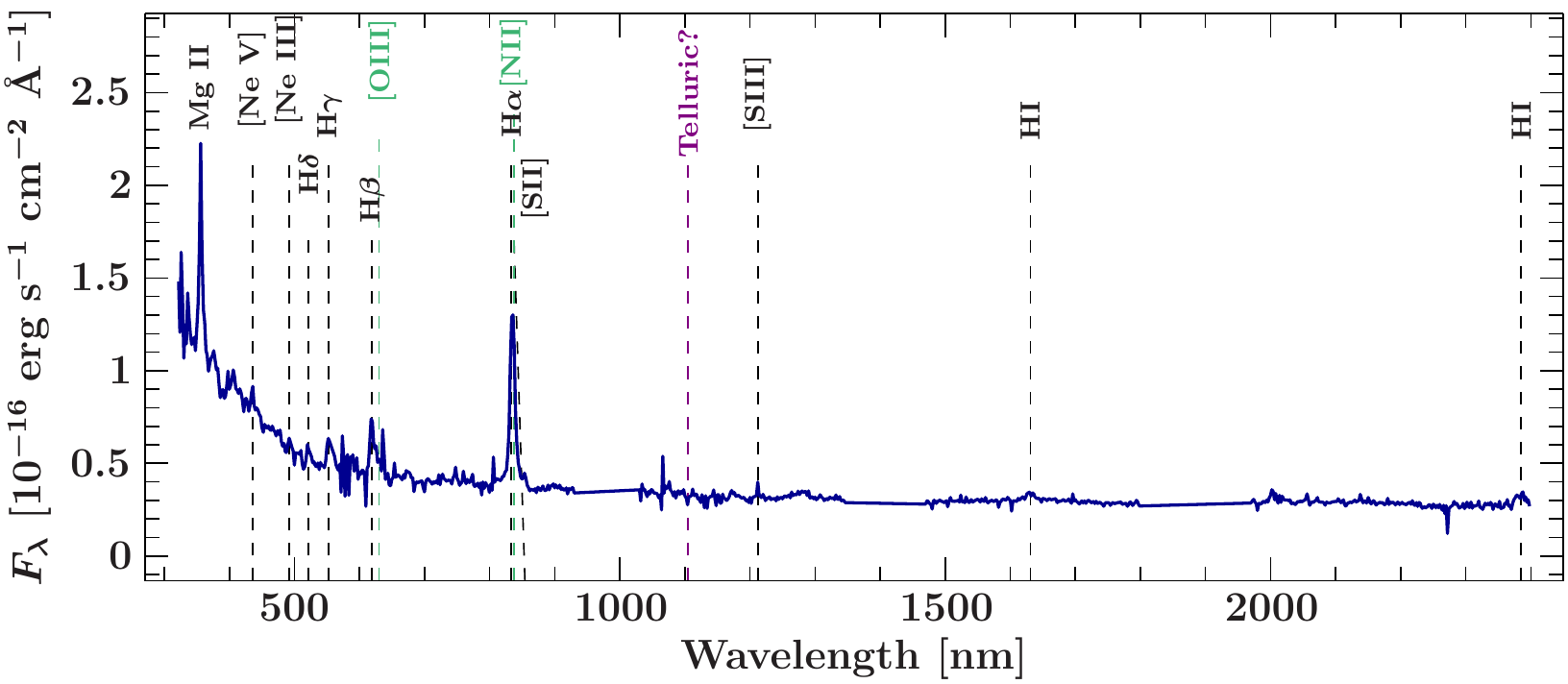}
    \caption{X-Shooter spectrum of 1WGA\,J2229.4$-$2018.}
    \label{fig:xshooter_spectrum}
\end{figure*}
Data were reduced using the ESO X-shooter pipeline~\citep{Goldoni2006,
  Modigliani2010} (v.2.9.3), producing a background-subtracted,
wavelength-calibrated spectrum. The extracted 1D spectrum was flux
calibrated with the X-shooter pipeline using a response function
produced by observing the white dwarf standard LTT 3218 (R.A. 08h 41m
32s.43, Dec. -32$^\circ$ 56$^\prime$ 32.9$^{\prime\prime}$, J2000)
during the same night.  
The spectrum was corrected from telluric absorption lines using the
Molecfit package \citep{Smette2015b}, and the flux was dereddened using
\citep{Cardelli1989} with A$_V$=0.1358 mag and R$_V$=3.1. 

The reduced 1D spectrum is shown in Figure
\ref{fig:xshooter_spectrum}. It is dominated by strong MgII (279.8\,nm
- rest frame) and Balmer H$\alpha$, $\beta$, $\gamma$, $\delta$
emission lines that allow to derive a source redshift of $z=0.27$.
Forbidden lines of [O\,III] (495.9\,nm; 500.7\,nm), [Ne\,V]
(342.6\,nm), [Ne\,III] (386.9\,nm), [N\,II] (658.3\,nm), [S\,II]
(671.6\,nm) and [S\,III] (953.1\,nm) are also present. While HI and
MgII broad allowed emission lines presumably form within the accretion
disk or close to it, narrow forbidden lines of neon, oxygen and sulfur
are supposed to come from lower density regions further away from the
supermassive black hole. 
The X-shooter rest-frame spectrum of 1WGA\,J2229.4-2018 matches with
the optical spectrum of a Seyfert 1.2 AGN with [O\,III] lines weaker
than the H$\alpha$ one. The $\nu F_\nu$ optical/near-infrared spectral
energy distribution of 1WGA\,J2229.4-2018 displays a deep trough around
10$^{14.8}$ Hz which probably indicates the transition between the dusty
torus and the disk contributions strengthening the classification of
the source as a Seyfert 1.2 AGN.

Fig.~\ref{fig:xshooter_spectrum} shows our spectrum of
1WGA\,J2229.4$-$2018. We derive a source redshift of $z = 0.27$ using
the emission lines of Mg II, Ne V, H$\beta$, and H$\alpha$ and we use
this number to calculate the absolute luminosity of the source and its
SED.
Based on the optical spectrum the source can be identified as Seyfert
type I AGN.

\subsection{ATCA Observation}
\label{sub:obs:atca}
Following the detection of the neutrino candidate \icnu, we requested
radio observations with the Australia Telescope Compact Array (ATCA;
under project code CX433) targeting the four X-ray sources found by
\swift/XRT within the neutrino location error region. The ATCA
observations were carried out on 2019 April 21 and 22. On the first
night ATCA observed X-ray sources 1, 3, and 4, while on the 
second night it targeted X-ray sources 1 and 2. The observations were
recorded simultaneously at central frequencies of 5.5 and 9\,GHz, with
2\,GHz of bandwidth at each frequency. We used PKS 1934$-$638 for
bandpass and flux calibration, while the nearby source J2203$-$188 was
used for phase calibration. 
The data were edited, calibrated, and imaged following standard
procedures within the Common Astronomy Software Application
\citep[CASA, version 5.1.0; ][]{McMullin2007}. Imaging was done using
a Briggs robust parameter of 2 to maximise sensitivity. Since X-ray
source \#1 was observed on both days, the two observations were
combined to maximise sensitivity. 

\begin{figure*}
    \centering
\includegraphics[width=0.95\textwidth]{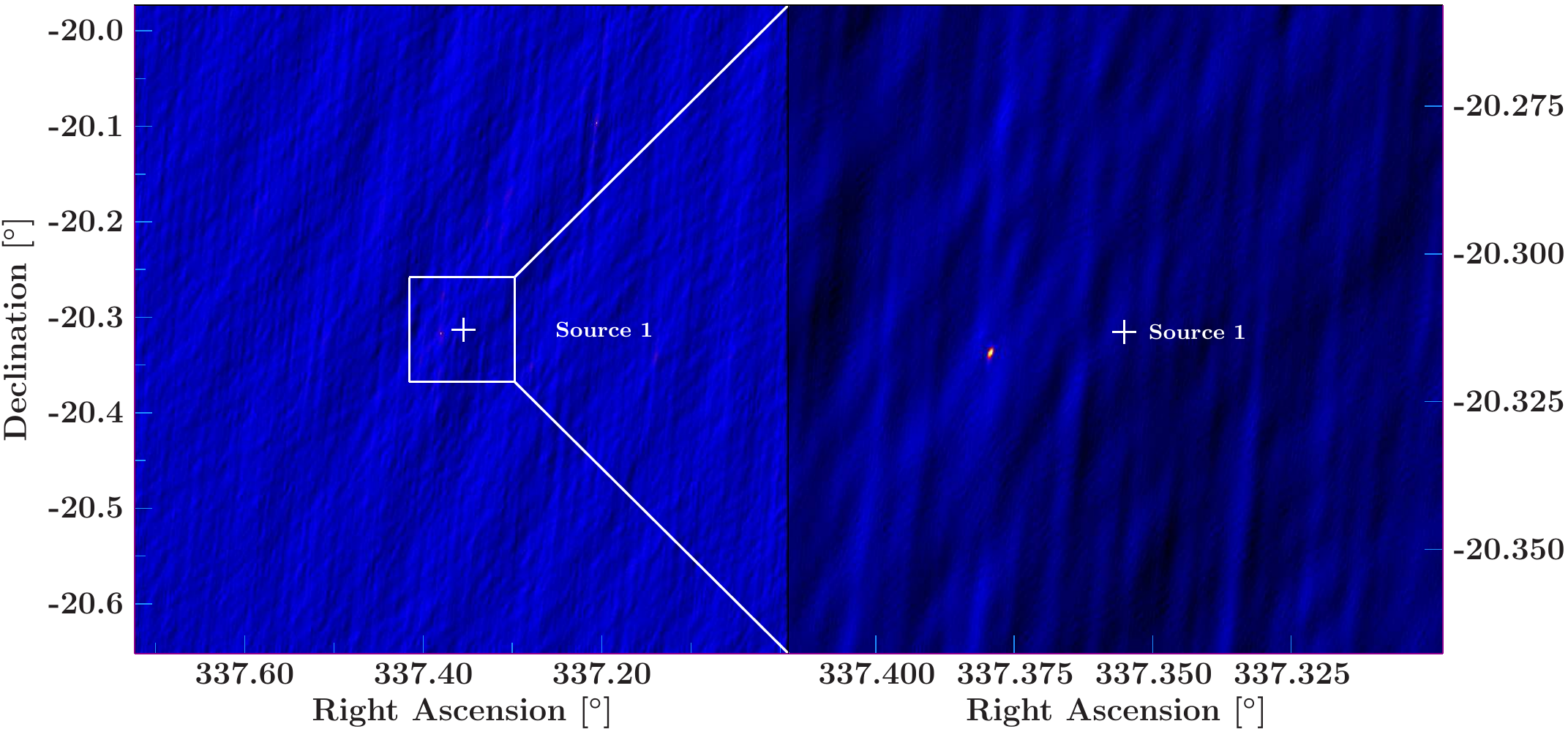}
    \caption{Stacked ATCA image near source 1. No radio source is
      detected that corresponds to the X-ray source. A radio source -
      likely a previously unknown AGN - is visible to the east of Source 1, at a
      distance of $2.52^{\prime}$.}
    \label{fig:atca_src1}
\end{figure*}

\begin{figure}
    \centering
\includegraphics[width=0.95\columnwidth]{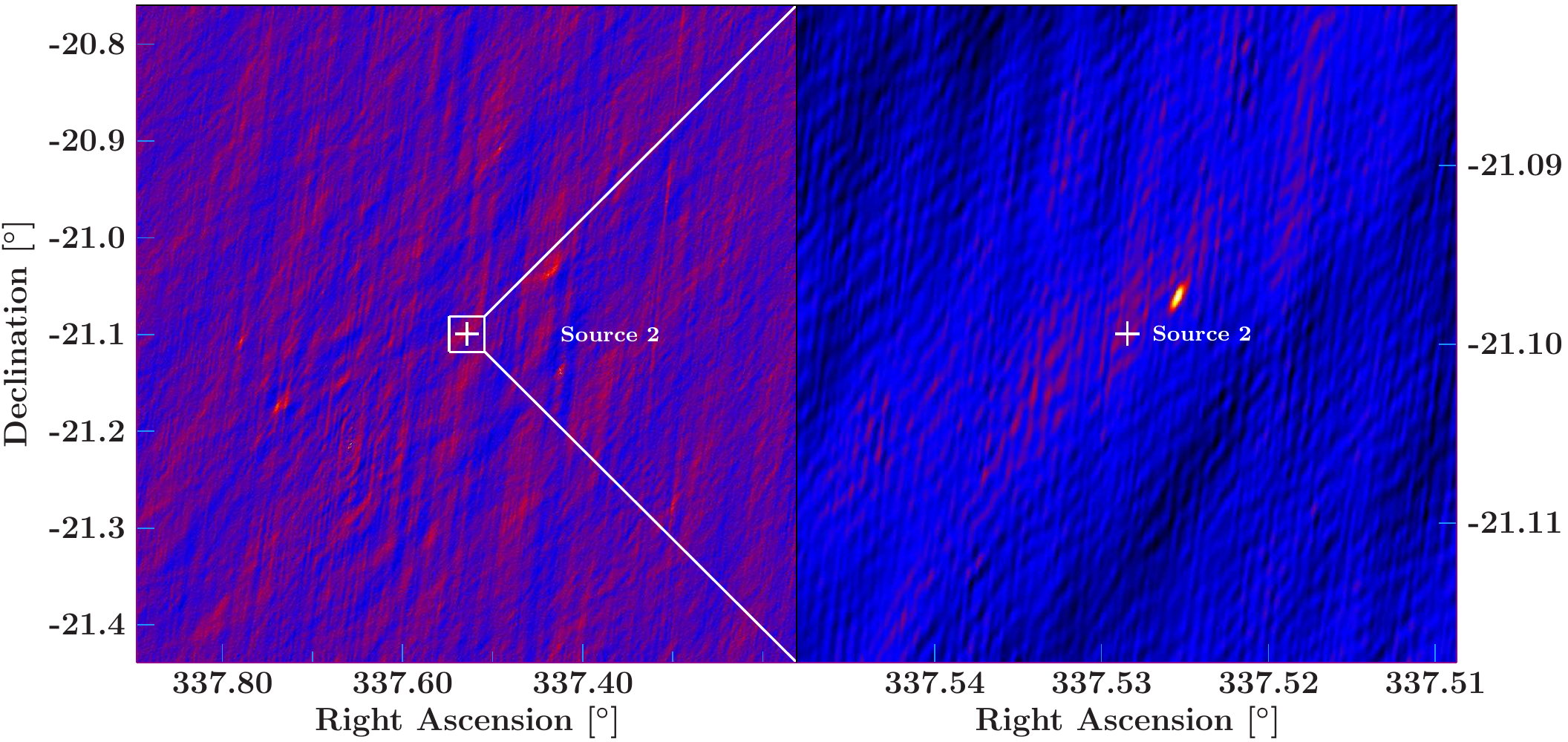}
    \caption{Stacked ATCA image near source \#2, a radio source is
      visible to the north-west. It is unclear whether it is connected
    to the X-ray detection.}
    \label{fig:atca_src2}
\end{figure}

No radio counterpart was detected at any of the X-ray source
positions. Upper limits were determined by stacking the 5.5 and
9\,GHz data and taking three times the measured rms over the source
position. They were extracted from regions of 30\arcsec centred on
the source positions. The flux densities of nearby sources are
reported as the peak pixel flux density. To determine the position of
these nearby radio sources, we fit for point sources in the image
plane. The resulting values are given in Table~\ref{tab-atca}.
\begin{table}\centering
  \caption{The ATCA flux upper limits and the
    corresponding coordinates. The ID gives the \swift\ /XRT source ID.
    Right ascension and declination give the centre of the region that
    was used for determining the upper limit. The coordinates for
    source \# 2 are offset due to flux from a nearby source. The uncertainties
    on the right ascension and declination are 0.2$^{\prime\prime}$
    and 0.5$^{\prime\prime}$, respectively.
  }
  \label{tab-atca}
  \begin{tabular}{ccccc}
    ID & RA & Dec & Flux UL \\
    &  [J2000] & [J2000] & [$\mu$Jy/beam]\\
    \hline
    1 & 337.35513 & $-$20.31324 & 12.36 \\
    2 & 337.53187 & $-$21.10525 & 27.62\\
    3 & 338.02441 & $-$21.04247 & 46.72\\
    \end{tabular}
\end{table}
ATCA finds no source consistent with the X-ray detection of source 1
(see Fig.~\ref{fig:atca_src1}); it only finds a source offset from the
X-ray position. Source 2 shows a possible ATCA counterpart with a
faint jet feature in the stacked image (Fig.~\ref{fig:atca_src2}).
Source 3 shows a very faint source south of the X-ray coordinates,
which is likely unrelated to the X-ray source. A brighter AGN is
visible, east of the coordinates.  

\begin{figure}
    \centering
\includegraphics[width=0.95\columnwidth]{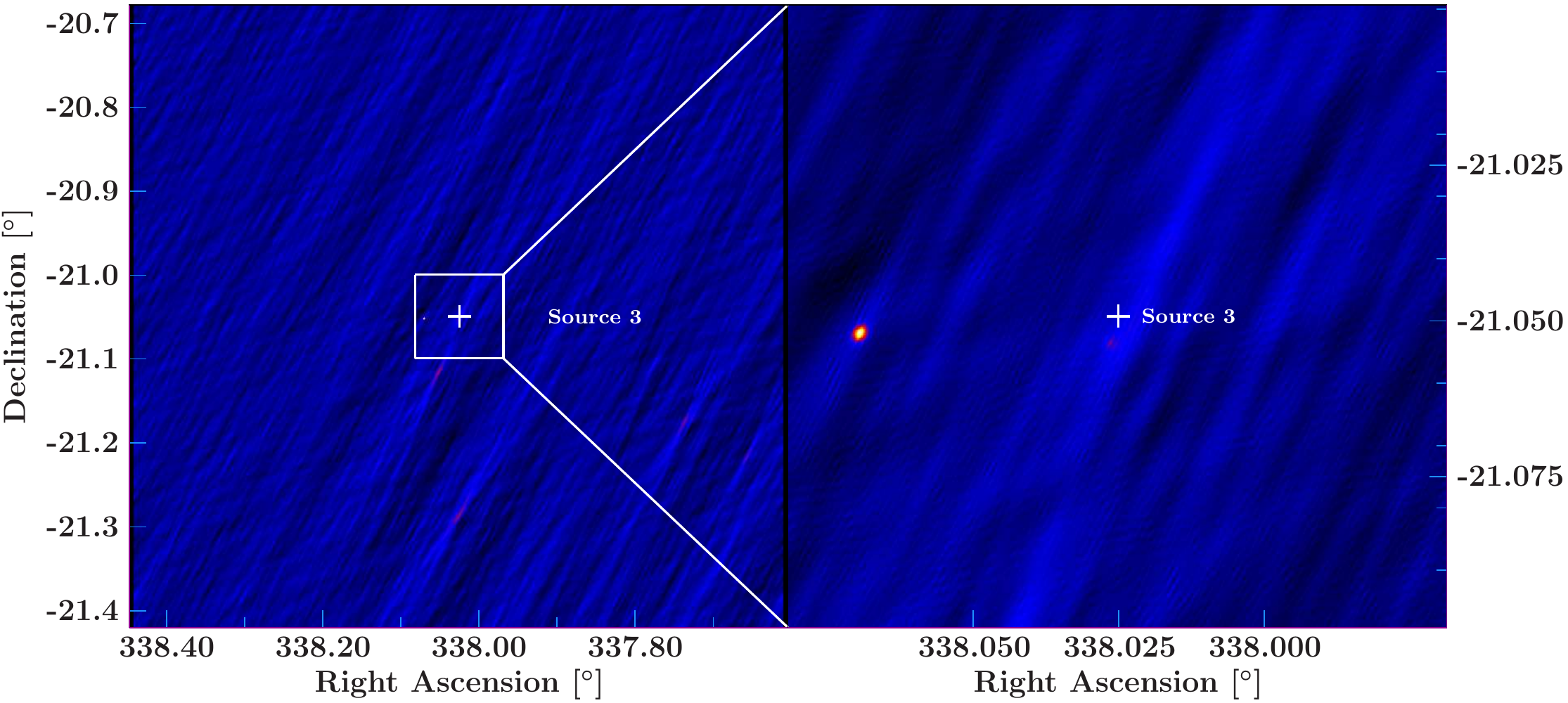}
    \caption{Stacked ATCA image near source 3. No bright counterparts
      is visible.}
    \label{fig:atca_src3}
\end{figure}

The derived flux upper limit for source \#1 is shown in the
multiwavelength SED (Fig.~\ref{fig-sed}).

\subsection{NuSTAR observations}{}

Following the detection of \icnu\ we requested target of opportunity
observations using the \nustar\ X-ray satellite
\citep{2013ApJ...770..103H} to search for hard X-ray sources
coincident with the neutrino event.  

\begin{figure*}
    \centering
    \includegraphics[width=0.7\textwidth]{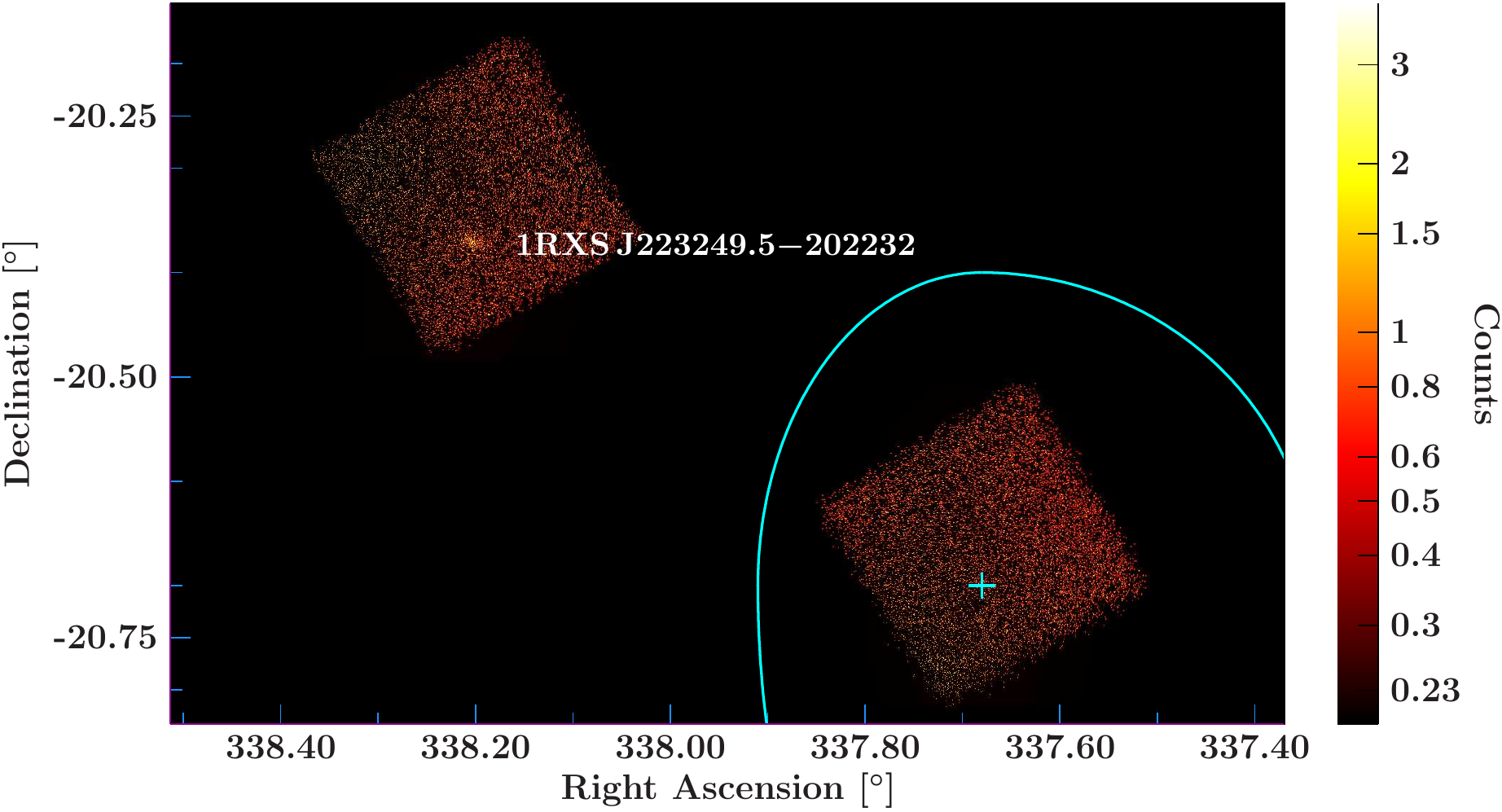}
    \caption{\nustar\ follow-up observations of the region around
      IceCube-190331A. The image shows the two \nustar\ pointings,
      with a strong detection of 1RXS J223249.5-202232. The cyan
      marker and egg-shape indicate the IceCube best fit and 90\%
      containment region for the neutrino event, respectively.} 
    \label{fig:nustar}
\end{figure*}

Two observations were performed: the first (ObsID 90502615001, started
on 2019 April 2 UT) targeting the best-fit neutrino position, and the
second (ObsID 90502616001, 2019 April 3) on the nearby \fermi/LAT
source 4FGL J2232.6-2023, associated with the hard X-ray source 1RXS
J223249.5-202232, a BL Lac object at a redshift of
0.386~\citep{2012AJ....143...64J}. A sky map containing data from both
exposures plus their relative locations with respect to the neutrino
event and \swift\ pointings is shown in Fig.\ref{fig:nustar}. 

Both focal plane modules (FPMs A and B) were used to collect data,
which were then processed using version 1.9.2 of the
\texttt{NuSTARDAS} software included in \texttt{HEASOFT} v6.27.2 and
analysed using \texttt{XSPEC} v12.11.0. 

\subsubsection{Observation of the \icnu\ position}

A total exposure of 5.5\,ks per FPM was collected at the
\icnu\ position. The observations did not reveal any new X-ray sources
and therefore we derive a flux upper limit at the best-fit neutrino
location. The observations from both FPMs were first combined and a
$3\sigma$ count rate upper limit was calculated using the
\texttt{uplimit} routine in \texttt{XIMAGE} which implements the
Bayesian approach of \cite{1991ApJ...374..344K}. This upper limit was
calculated for a circular region with a 30'' radius centred at the
best-fit neutrino location, but given the homogeneity of the field we
expect it to be illustrative of the entire \nustar\ exposure. The
count rate upper limit is 4.66 counts / ks, which corresponds to a
flux upper limit of $1.604\times 10^{-13}$ erg cm$^{-2}$ s$^{-1}$ in
the 3-10 keV range as calculated using the WebPIMMS
tool\footnote{\url{https://heasarc.gsfc.nasa.gov/cgi-bin/Tools/w3pimms/w3pimms.pl}}
for a photon index of $\Gamma = 2.0$. 

\subsubsection{Observation of 1RXS J223249.5-202232}

An exposure of 5.1 ks per FPM was obtained targeting 1RXS
J223249.5-202232. The spectral data from both FPMs were combined using
the \texttt{addspec} routine following the procedure recommended by
the \nustar\ science team and then grouped requiring at least 30
counts in each spectral bin. The resulting spectrum covers the 3-15
keV range with good statistics after bad channels are excluded. The
spectrum was then fit with an absorbed power-law model (\texttt{phabs}
$\times$ \texttt{powerlaw}, PL hereafter) where the absorption was
kept fixed at the Galactic HI column contribution of $3.33 \times
10^{20}$ cm$^{-2}$ obtained from
\cite{2016A&A...594A.116H}\footnote{\url{https://heasarc.gsfc.nasa.gov/cgi-bin/Tools/w3nh/w3nh.pl}}.  

The best-fit PL parameters (for the form $N(E) = N_0 \; E^{-\Gamma}$)
were a flux normalisation $N_0 = (1.88 \pm 0.76) \times 10^{-3}$
cm$^{-2}$ keV$^{-1}$ s$^{-1}$, at a normalisation energy of 1 keV, and
a photon index $\Gamma = 2.87 \pm 0.24$. These parameters yield a good
fit, with a $\chi^2$/dof = 6.33/7 ($p$-value of 0.502). A second fit
was attempted to test for intrinsic absorption at the source using the
\texttt{zphabs} model and the source redshift but this failed to
constrain the intrinsic absorption given the limited statistics of the
data set. 

Using the best-fit model we calculate a flux of $F =
(1.39^{+0.09}_{-0.27}) \times 10^{-12}$ erg cm$^{-2}$ s$^{-1}$ in the
2-10 keV range. This flux is significantly lower and has a softer
photon index than the values reported by \cite{2012AJ....143...64J}
based on hard-band \emph{XMM-Newton} EPIC observations obtained in
2008 in the same energy range ($F_{\mathrm{XMM}} = (5.05 \pm 0.04)
\times 10^{-12}$ erg cm$^{-2}$ and $\Gamma_{\mathrm{XMM}} = 2.06 \pm
0.03$).  

\subsubsection{Summary of \nustar\ results}

No new sources were identified in the \nustar\ observations near the
best-fit neutrino position. Similarly, the known source 1RXS
J223249.5-202232 was observed in a low flux state, and no evidence of
gamma-ray activity is visible in the online \emph{Fermi} All Sky
Variability Analysis
(FAVA)\footnote{\url{https://fermi.gsfc.nasa.gov/ssc/data/access/lat/FAVA/LightCurve.php?ra=338.1725&dec=-20.3909}},
nor was reported at the time of the neutrino alert
detection\footnote{\url{https://gcn.gsfc.nasa.gov/gcn3/24040.gcn3}}.  

We therefore claim no connection between this source and the
observation of the \icnu\ event. The \nustar\ pointings do not cover
the new \emph{Swift} sources so no constraints can be derived on their
hard X-ray fluxes from these observations.

\subsection{Helix Nebula}

We note that the planetary nebula NGC 7293, more commonly known as the
Helix Nebula, is detected on the western side of the UVOT mosaic,
within neutrino localisation area. The expected particle energies in a
planetary nebula, however, are too low to explain a TeV neutrino event. 
Planetary nebulae can have fast winds up to $\sim 1100\,$km\,s$^{-1}$
with a mass loss rate of $\dot{m}\sim
10^{-11}M_{\astrosun}$\,year$^{-1}$ \citep{Huarte2012}.
Using the kinetic luminosity of $L_k=\dot{m}\,v^2$, this yields the
following magnetic luminosity \citep{Blandford2000}
\begin{equation}
L_B = \epsilon_B \, L_k = 4\,\pi\, r^2\cdot v\cdot
\left(\dfrac{B^2}{8\,\pi}\right) = \dfrac{1}{2} r^2 \cdot v B^2\qquad ,
\end{equation}
with the magnetic energy fraction $\epsilon_B<1$, the magnetic field
$B$ and the radius of the moving wind $r$. 
After solving for $r\cdot B$ and using the previously calculated
kinetic luminosity, we can use it in
\begin{equation}
  E<e\cdot r \cdot B \cdot \dfrac{v}{c}\qquad ,
\end{equation}
using the Hillas condition \citep[v/c;][]{Hillas1984}, with the
elementary charge $e$ and the speed of light $c$, to yield 
\begin{equation}
  E < \sqrt{2 \dot{m} v}
\end{equation}
This yields a maximum particle energy of
\begin{equation}
E<0.41\,\mathrm{TeV}\qquad .
\end{equation}

The peak energy of neutrinos for E$^2$\,d$N$/d$E$ for incident protons with single
energy injection is about 3--5\% of the energy of the proton
\citep[e.g.,][]{Waxman1997,Kelner2006,2006ApJ...651L...5M}.
We would therefore not expect neutrinos above $\unsim 10$--$20$\,GeV
from the Helix nebula. For this reason, planetary nebulae have not
been considered as sources for high-energy IceCube neutrinos.

\section{Results \& Discussion}
\label{sec:disc}

We found four X-ray sources in the \swift/XRT observations. No new
sources, or evidence of X-ray activity in a known source, were
identified in the \nustar\ observations. We perform follow-up
observations on all of them, particularly the brightest three sources.
We  note that only source \#2 is strictly within the IceCube
90\% confidence uncertainty region.

We performed detailed follow-up observations of Source \# 1, the a
priori most likely counterpart based on source brightness. The data
has been gathered and collected in Fig.~\ref{fig-sed}. With no radio
detection and no \fermi/LAT detection, but a moderate X-ray luminosity
($L_{2-10\,\mathrm{keV}}=1.3^{+2.1}_{-0.9}\times10^{43}$\,erg\,s$^{-1}$),
we conclude that this source is a possible radio-quiet quasar. A
X-Shooter optical spectrum confirms that this object is a type 1
Seyfert galaxy. Neutrinos have been predicted from the cores of  AGN
in the 10--100\,TeV energy range \citep{Murase2019}. The flare
contribution may be subdominant while the core contribution can be
dominant in the bulk flux \citep{Murase2019}. This would change the
current picture that particle acceleration in AGN jets is the dominant
way to produce neutrinos from AGN \citep{bb,ic170922a}, and
potentially counterindicated by the high observed energy  of this
neutrino event.

Neither of the other X-ray sources show obvious counterparts in
\fermi/LAT or in ATCA observations. Radio sources are detected close
to source \#2 and source \#3, but are not obviously counterparts to
the XRT detections. As they have no known counterparts and the X-ray
spectrum cannot distinguish between relevant models, we cannot
speculate on what source type they are. They could be AGN, or Galactic
X-ray sources, such as compact binary objects. However,
IceCube-190331A was detected at a Galactic latitude of $-57\fdg31$,
which shows that a Galactic origin is unlikely.

\subsection{X-ray coincidences}

Here, we examine whether observing 1 X-ray sources within and 3 other
near the neutrino uncertainty is noteworthy. Given the uncertainties
in the right ascension and the declination ra$_{+}$, ra$_{-}$,
dec$_+$, and dec$_-$, the equation for the area of the uncertainty
region is given by
\begin{equation}
  A_\mathrm{unc} = \dfrac{\pi}{2}\cdot \left(\dfrac{1}{2}~
\mathrm{dec}_+ \cdot (\mathrm{ra}_{+}+\mathrm{ra}_{-})\, + \,
\dfrac{1}{2}~ \mathrm{dec}_- \cdot (\mathrm{ra}_{+}+\mathrm{ra}_{-})/2 \right) 
\end{equation}
which yields 0.35\,deg$^2$ for IceCube-190331A.  Currently, there is
no deep full scan of the sky available in the \textsl{Swift} X-ray
band. A full catalog would give us a precise estimate of the number of
X-ray sources expected in this region. As an approximation, we use the
ROSAT catalogue as a comparison tool \citep{ROSAT}. The 2RXS catalogue
includes 135,118 sources. We select sources above Galactic latitudes
of $\pm 10^{\circ}$ to give an estimate of the number of extragalactic
sources, which yields 117,094 sources. The surface area of the night
sky is 41253 deg$^2$. The Galactic plane area that we exclude is given
by $A_\mathrm{excl} = 2\cdot\pi\cdot r\cdot h = 7200^\circ$, with the
radius $r=57.3^\circ$ and the height $h=20^\circ$. This yields a total
area for the ROSAT sources of $A_\mathrm{total}=34053$\,deg$^2$. For
the IceCube neutrino event, we'd therefore expect $\sim 1.2$ ROSAT
sources within  the uncertainty region of 0.35\,deg$^2$. 
While we do find exactly one X-ray source within the uncertainty
region, it is not a ROSAT source. This is acceptable considering the
low number statistics, and therefore does not point towards an
association of the neutrino event and the X-ray source. Given that
\swift/XRT has a wider energy range and more sensitivity, and that the
observations were pointed, it is not surprising to find more X-ray
sources in the near vicinty.


\section{Conclusions}

We have performed \swift/XRT, \swift/UVOT and \nustar\ follow-up
observations of the IceCube neutrino alert IceCube-190331A. This event
is important as it has a high likelihood of being astrophysical in
origin (higher than IceCube-170922A). We find four X-ray sources in
the tiled \swift/XRT mosaic observations, with two having high
detection significance, while no new sources, or activity in a known
one, were identified in the \nustar\ observations.

The brightest \swift/XRT source (\#1) is consistent with
\mass . Due to its known optical counterpart and its X-ray brightness
it seemed to be the most likely source of neutrinos. A high X-ray
brightness is required to explain the expected electromagnetic
emission from secondary cascades of hadronic particles. The lack of
$\gamma$-ray emission from the source is not fully consistent with
this picture. The inconsistency may be explained if the high densities
required for neutrino production in the source cause $\gamma$ pair
production of the high-energy photons \citep{Zhang1997}. 
Additionally, we performed follow-up observations of source \#1 with
X-Shooter and ATCA. The source is not detected in ATCA with strong
constraints on the radio flux, and it is not detected by \fermi/LAT.
Given the radio-quietness, the low $\gamma$-ray flux and 
the X-ray detection, the source is either not a blazar or a very
faint/distant one. Our X-Shooter spectra has confirmed that source
\#1 is a type 1 Seyfert galaxy. ATCA follow-up of sources \#2 and \#3
show possible radio counterparts near the X-ray position. However,
neither of them has been detected by \fermi/LAT and they are not in
any catalogue. We therefore conclude that there is no blazar counterpart
and no other obvious high-energy source counterparts as such, and its
likely astrophysical origin remains a mystery. The neutrino
localisation region was serendiptidously located near the highest
sensitivity location of the field-of-view of \swift /BAT. This 
constrains and rules out a bright GRB at the time and location of the
neutrino. For future IceCube events at high signalness (high
probability of being astrophysical in origin), a more rapid
multiwavelength response with quasi-simultaneous data will help
greatly in identifying the sources of high-energy neutrino alerts.

\section*{Acknowledgements}
We thank the anonymous referee for useful comments that have improved
the manuscript.
We thank R. Ciardullo for useful discussions and feedback on the draft.
F. K. was supported as an Eberly Research Fellow by the Eberly College
of Science at the Pennsylvania State University.
P. A. E. acknowledges UKSA support.
P. M. acknowledges support from the Eberly Foundation.
M. S. is supported by NSF awards PHY-1914579 and PHY-1913607.
We thank J.E.~Davis for the development of the \texttt{slxfig} module
that has been used to prepare the figures in this work. This research
has made use of a collection of ISIS scripts provided by the Dr. Karl
Remeis-Observatory, Bamberg, Germany at
\url{http://www.sternwarte.uni-erlangen.de/isis/}. This research has
made use of the NASA/IPAC Infrared Science Archive, which is funded by
the National Aeronautics and Space Administration and operated by the
California Institute of Technology.
This publication makes use of data products from the Two Micron All
Sky Survey, which is a joint project of the University of
Massachusetts and the Infrared Processing and Analysis
Center/California Institute of Technology, funded by the National
Aeronautics and Space Administration and the National Science
Foundation. 




\bibliographystyle{mnras}
\bibliography{ic190331a}

\section*{Data availability}
Data from \textsl{Swift} and \textsl{NuSTAR} data are publicly
available on HEASARC (\url{https://heasarc.gsfc.nasa.gov/}). 2MASS and
WISE data are available publicly at
\url{https://irsa.ipac.caltech.edu/frontpage/}. Raw ATCA data are
available at \url{https://atoa.atnf.csiro.au/query.jsp}. X-Shooter data are available on request.




\bsp	
\label{lastpage}
\end{document}